\documentclass[letterpaper,10pt]{aa}
\usepackage{graphics}
\usepackage{natbib}
\begin{document}
\newcommand{\Msun}{M$_{\sun}$\ }
\newcommand{\Lsun}{L$_{\sun}$\ }
\newcommand{\Mstar}{M$_{*}$\ }
\newcommand{\Lstar}{L$_{*}$\ }
\newcommand{\aar}{A{\&}A Rev.\ }
\newcommand{\aas}{A{\&}AS\ }
\newcommand{\annrev}{ARA{\&}A\ }

\title{Proper Motion Studies of Outflows from Classical T Tauri Stars}
\author{F. McGroarty, \inst{1,} \inst{2}
T. P. Ray \inst {2} \and D. Froebrich \inst {3}}
\offprints{F. McGroarty, fiona.mcgroarty@nuim.ie}
\institute{Department of Experimental Physics, National University of Ireland Maynooth, Maynooth,
Co. Kildare, Ireland
\and Dublin Institute for Advanced Studies, 5 Merrion Square,
Dublin 2, Ireland
\and Centre for Astrophysics \& Planetary Science, School of Physical Sciences, University of Kent, Canterbury CT2 7NH, UK}
\date{Received date ;accepted date}

\abstract{
In a previous paper \citep{McGroarty04b} we examined the environment of a number of evolved low\,-\,mass young stars, i.e. 
Classical T Tauri Stars, to see if they are capable of driving parsec\,-\,scale outflows. These stars - CW Tau, DG Tau, DO 
Tau, HV Tau C and RW Aur - were previously known to drive only ``micro\,-\,jets'' or small\,-\,scale outflows of  
$\leq$\ 1\arcmin\  or 0.04 pc at the distance of the Taurus\,-\,Auriga Cloud. We found that they drive outflows of 0.5pc -- 
1pc, based on the morphology and alignment of newly discovered and previously known HH objects with these sources and their 
``micro\,-\,jets''. Here, we use a cross\,-\,correlation method to determine the proper motions of the HH objects in these 
five outflows (HH\,220, HH\,229, HH\,702, HH\,705 and HH\,826 -- HH\,835) which in turn allows us to confirm their driving 
sources. Moreover, the tangential velocities of HH objects at large distances from their origin are currently poorly known so
these proper motions will allow us to determine how velocities evolve with distance from their source. We find tangential 
velocities of typically 200 kms$^{-1}$ for the more distant objects in these outflows. Surprisingly, we find similar 
tangential velocities for the ``micro\,-\,jets'' that are currently being ejected from these sources. This leads us to 
suggest that either the outflow velocity was much higher 10$^3$ years ago when the more distant objects were ejected and 
that these objects have decelerated to their current velocity or that the outflow velocity at the source has remained 
approximately constant and the more distant objects have not undergone significant deceleration due to interactions with 
the ambient medium. Numerical simulations are needed before we can decide between these scenarios.

}
\maketitle

\section{Introduction}
Herbig\,-\,Haro (HH) objects are the optical emission line tracers of outflows from young stars. Although the exact 
mechanism by which outflows are ejected from a young forming star are still unclear, it is accepted that they are 
intrinsically linked to accretion \citep{Cabrit90,Hartigan95}. Close to the source these HH objects are usually seen as well
collimated jets, however further out they separate into discrete knots or diffuse regions, in some cases taking the form of 
a bow shock -- the objects in the HH\,34 outflow is a typical example of such as structure \citep{Devine97}. Initially these
HH outflows were assumed to be a fraction of a parsec in length, with only a few exceptions (e.g. Ray, 1987). However the 
past decade has revealed many parsec\,-\,scale outflows \citep{Bally97,Reipurth01}, which are often comparable in size 
to their parent cloud.\\  

The majority of outflows are found to be driven by low\,-\,mass young stars, which can be classified by the amount of excess 
infrared emission in their spectral energy distribution.  
This excess is due to circumstellar dust and is very strong in the young, embedded Class I sources and is almost 
non-existent in the most evolved Class III sources \citep{Lada84,Lada87}. Parsec\,-\,scale optical outflows are generally 
found to be driven by low\,-\,mass Class I sources, however a number of intermediate\,-\,mass sources have recently been 
observed to drive large\,-\,scale outflows \citep{McGroarty04a}.\\ 

In \cite{McGroarty04b} (hereafter referred to as {\bf MR04}) we examined a number of more evolved, Class II low\,-\,mass 
sources (i.e. Classical T Tauri Stars - CTTSs) and found they also drive large\,-\,scale outflows. The sources observed 
in MR04 (CW Tau, DG Tau, DO Tau, HV Tau C and RW Aur) were previously only known to drive ``micro\,-\,jets'' or 
small\,-\,scale outflows of $\leq$ 1\arcmin. This corresponds to 0.04 pc at the distance (140pc) to the Taurus\,-\,Auriga 
Cloud \citep{Elias78, Wichmann98}. Our observations in MR04 show them to actually drive outflows of $\sim$ 0.5 - 1 pc. 
Although these outflows are not as large, or as spectacular, as those from younger sources, they are less embedded so are 
more easily observed.\\ 

\cite{Hartigan95} have shown that mass outflow rates are directly correlated with mass accretion rates. The stellar envelope
dissipates over time as material is accreted onto the protostar thus accretion rates should decrease with time 
\citep{Ward-Thompson02}. \cite{Bontemps96} show observational evidence for this with a decline in molecular outflow 
strength from Class 0 to Class I Young Stellar Objects (YSOs). We can infer a further decline in Class II YSOs. Despite this
CTTSs are still capable of driving collimated atomic/ionized large-scale outflows (MR04). We see that their morphologies are
similar to those from Class I sources, but how do the lower accretion rates affect their kinematics? Lower accretion rates
translate into lower mass loss rates and hence a decrease in the mechanical luminosity. However, as the average ambient 
density is lower in a Class II environment the outflow velocity may not be reduced much by interaction with its surroundings.
Determining velocities for the distant HH objects in these CTTS outflows is thus fundamental to discovering the effect that 
the lower accretion rates of CTTSs and the lower parent cloud density will have on outflows driven by such sources.\\

As a continuation of the MR04 study, here we use multi-epoch observations to determine proper motions and hence tangential 
velocities for the HH objects previously discovered from our CTTS sources. This will allow us to test whether the driving 
sources determined in MR04, largely based on outflow morphology and alignment, are correct. Additionally, the tangential 
velocities of HH objects at relatively large distances from their source are poorly known. In particular, how these 
velocities evolve with distance from the source is not well understood. Both of these issues are addressed here in light of 
the results of the proper motion studies.\\ 

Dynamical timescales were estimated for these outflows in MR04 based on the assumption that the most distant objects are 
moving with a tangential velocity of 50 kms$^{-1}$, which is a lower limit for the velocity needed to induce optically 
visible shocks. This is a conservative estimate and the more accurate tangential velocities obtained will allow more 
realistic dynamical timescales to be calculated. In any event, however, it is important to note that these timescales are 
much less than the age of the outflow as estimated from evolutionary tracks, and only represent their {\it optically visible}
portion. There is little doubt that these outflows have blown out of their parent cloud (MR04).\\

Details about the observations and proper motion method used are given in \S\ref{sec-observations} and 
\S\ref{sec-propermotions} respectively. The results of this study are reported in \S\ref{sec-results4} and the implications 
of these results are discussed in \S\ref{sec-discussion} with our conclusions being presented in \S\ref{sec-conclusion}.

\section{Observations}
\label{sec-observations}
Using multi-epoch observations the velocity and direction of motion are determined for many of the known HH objects in the 
vicinity of CW Tau, DG Tau, DO Tau, HV Tau C and RW Aur. The observations used here were taken on a number of different 
observing runs using the Wide Field Camera on the Isaac Newton Telescope, La Palma (Canary Islands). The same CCDs and hence
the same angular resolution is used for all observations -- 1 pixel projects to 0\farcs33 square on the sky.
The observing runs occurred in November 1999 (epoch 1999.91), February 2001 (epoch 2001.12), December 2002 (epoch 2002.98) 
and finally November 2003 (epoch 2003.91), however not all regions were targeted during each run. Narrow band filters were 
used to observe the HH objects, both [SII]($\lambda_c$ = 672.5nm, $\Delta\lambda$(FWHM) = 8.0nm) and H$\alpha$($\lambda_c$ = 
656.8nm, $\Delta\lambda$(FWHM) = 9.5nm) are used here, but again not all regions were observed using both of these filters 
in each run. Table\ \ref{ObsLog} lists the observations for each region. Exposure times for the narrow band images were  
typically 30 minutes. The data was reduced using standard IRAF procedures for bias subtraction and flat fielding.\\ 

\section{Proper Motion Method}
\label{sec-propermotions}

A cross\,-\,correlation method was devised to determine the proper motions of the HH objects in the vicinity of our five CTTS
sources. The images from two different epochs are initially aligned using star positions obtained from the SExtractor 
software \citep{Bertin96}. Our method then maps one of the images onto the other using a polynomial fit (up to third order) 
of the star positions and the IRAF task {\em geomap}. This procedure ensures the correction of any systematic effects such 
as stretching, rotation, etc. of the image.\\  

The smallest time difference between epochs in Table\ \ref{ObsLog} is 0.93 years (epoch 2003.91 - epoch 2002.98). A 
bright, compact HH knot 
that has a tangential velocity of 50 kms$^{-1}$ at a distance of 140pc will appear to shift 0.21 pixels with respect to the 
background stars in this time. Our cross correlation method samples to a spatial frequency of 0.1 pixels, so even such low 
velocities could, in principle, be detected.\\

\begin{table}
\begin{tabular}{llccc}
\hline \hline
Region       &Line      &$1^{\rm st}$ Epoch &$2^{\rm nd}$ Epoch &$3^{\rm rd}$ Epoch \\ \hline
DG Tau       &[SII]             &1999.91       &2001.12        &2003.91            \\ 
             &H$\alpha$         &2001.12       &2003.91        &                   \\ 
CW Tau       &[SII]             &2001.12       &2002.98        &2003.91            \\ 
             &H$\alpha$         &2001.12       &2002.98        &2003.91            \\ 
DO Tau       &[SII]             &2001.12       &2002.98        &2003.91            \\ 
             &H$\alpha$         &2001.12       &2002.98        &2003.91            \\ 
HV Tau C     &[SII]             &2001.12       &2002.98        &2003.91            \\ 
             &H$\alpha$         &2001.12       &2002.98        &2003.91            \\ 
RW Aur       &[SII]             &2001.12       &2003.91        &                   \\ 
             &H$\alpha$         &2001.12       &2003.91        &                   \\  \hline
\end{tabular}
\caption{Log of observations used for each source in the 
proper motions study.}
\label{ObsLog}
\end{table}

\begin{figure}    
\resizebox{\hsize}{!}{\includegraphics{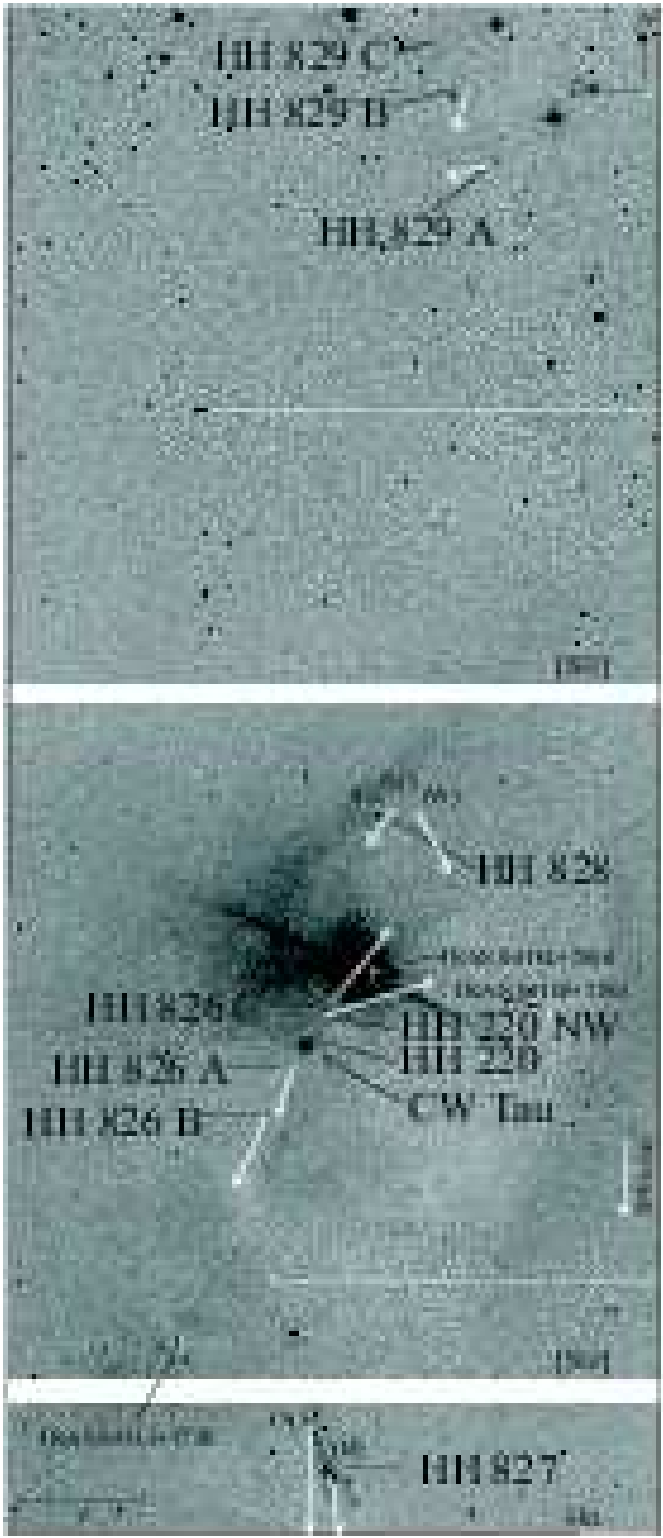}}
\caption
{Proper motions of HH\,826 -- HH\,829 in the vicinity of CW Tau. The direction of motion of each HH object is represented by 
white arrows. The relative length of these arrows denotes the relative velocity of the object. The redshifted (HH\,220) jet 
is also marked above, however it is too faint for proper motion studies.
These vectors show that HH\,826 and HH\,827 are driven by CW Tau, assuming precession of the outflow over time. 
However this study also shows that HH\,828 and HH\,829 are not driven by CW Tau as previously suggested. Candidate sources 
for these objects are suggested in the text.} 
\label{cwtau_flow_pms}        
\end{figure}

However, there are two main sources of error in these measurements. The first is due to the accuracy of the alignment of the
images. This is typically 0.16 -- 0.4 pixels (i.e. $\pm$(0.08 -- 0.2) pixels) in both x and y directions, generating errors 
of about 18 -- 56 kms$^{-1}$ or $\pm$ (9 -- 28) kms$^{-1}$. 
This is determined by using all combinations of epochs listed in Table\ \ref{ObsLog}, so is valid for all epoch 
separation timescales.
All errors in velocity and direction are quoted at 1$\sigma$.
However the errors in the central frame of the CW Tau region (i.e. HH\,826, HH\,220\,NW and HH\,828, see 
Fig.\ \ref{cwtau_flow_pms}) are much higher than this. There are very few stars in this field and consequently the alignment
is poor. Alignment errors for these objects are about 1 pixel in x and y i.e. $\sim$ 110 kms$^{-1}$ ($\pm$ $\sim$ 55 
kms$^{-1}$). 
The second source of error is in the position of the HH object and is due to its signal\,-\,to\,-\,noise ratio, PSF (point 
spread function) and extent. This seems to be the limiting error for most of the objects observed here, especially the faint
objects as one would expect. Typically these errors are about 0.2 -- 0.5 pixels (i.e. $\pm$ (0.1 -- 0.25) pixels) in x and y,
corresponding to $\sim$ 20 -- 55 kms$^{-1}$ or $\pm$ (10 - 28) kms$^{-1}$.  
The most noticeable exception are the very faint HH\,830 knots (in the vicinity of DG Tau) which in [SII] have errors of 
about 1.1 pixels, i.e. $\sim$ 120 kms$^{-1}$ ($\pm$ $\sim$ 60 kms$^{-1}$)  
and in H$\alpha$ have errors of $\sim$ 0.7 pixels in x and y i.e. $\sim$ 80 kms$^{-1}$ ($\pm$ $\sim$40 kms$^{-1}$). 
The errors in velocity for each frame/image as explained above are given in the caption for each table where results are 
stated. The errors in direction of motion are given separately for each object, as the error in direction depends on the 
velocity -- see Tables\ \ref{pms_cwtau} to \ref{pms_rwaur}.\\

A common problem with proper motion measurements can be that only the global motion of a HH object is obtained. Here, 
however, we have managed to get proper motions for individual knots in some of the larger HH objects, e.g. HH\,827, HH\,702, 
HH\,831 and HH\,705. In some cases it is advantageous to look at the global motion of a HH complex to ascertain a possible 
driving source and then to examine the individual motions within it. This has been done for HH\,827 (\S\ref{sec-CWTAU}), 
HH\,702 (\S\ref{sec-DGTAU}) and HH\,705 and HH\,831 (\S\ref{sec-DOHVTAU}).\\

One of the main causes of false proper motions is photometric variability. Occasionally, the majority of knots in a large 
HH complex will be moving in one direction, while an individual knot may appear to move in a completely different direction.
In this case, the different motion may be assigned to a change in relative brightness within that knot rather than 
physical motion. 
That said, the cooling time is typically much longer than the timespan between our measurements so such false measurements 
are usually not a problem.

\section{Results}
\label{sec-results4}
A detailed introduction to the CTTSs DG Tau, CW Tau, DO Tau, HV Tau C and RW Aur and their associated 
``micro\,-\,jets''/small\,-\,scale outflows is given in MR04. The newly discovered objects (HH\,826 -- HH\,835, see 
Table\ \ref{HHPositions}) that may be 
extensions to these outflows are also presented there. Here, a brief summary is given for each outflow before the proper 
motion results are discussed.

\begin{table}
\begin{tabular}{llllccc}
\hline \hline
Object       &Suggested             &Angular              &P.A.        \\ 
             &Source$^a$           &Separation           &   \\ \hline
HH\,826\,A   &CW Tau             &0\farcm37            &153\degr\   \\          
HH\,826\,B   &CW Tau             &1\farcm27            &153\degr\   \\
HH\,827      &CW Tau             &6\farcm2             &184\degr\   \\
HH\,826\,C   &CW Tau             &0\farcm77            &326\degr\ \\
HH\,828\,E   &\ \ \ \ \ \  ?$^b$                                      \\
HH\,828\,M   &\ \ \ \ \ \  ?$^b$                                      \\
HH\,828\,W   &\ \ \ \ \ \  ?$^b$                                      \\
HH\,829\,A   &CW Tau             &14\farcm9            &348\degr\     \\
HH\,829\,B   &CW Tau             &16\farcm12           &351\degr\   \\
HH\,829\,C   &CW Tau             &16\farcm8            &353\degr\    \\

HH\,830\,A   &DG Tau             &9\farcm6             & 48\degr\    \\
HH\,830\,B   &DG Tau             &11\farcm3            & 48\degr\    \\
HH\,830\,C   &DG Tau             &14\farcm3            & 48\degr\  \\

HH\,831\,A   &DO Tau             &10\farcm8            & 74\degr\  \\
HH\,831\,B   &DO Tau              &11$'$               & 74\degr\   \\
HH\,832      &DO Tau             &7\farcm7            & 78\degr\    \\
HH\,833      &HV Tau C           &4\farcm6            & 25\degr\    \\
HH\,834      &\ \ \ \ \ \   ?$^c$                                    \\

HH\,835      &RW Aur             &5\farcm37           &310\degr\    \\ \hline

\end{tabular}
\caption{Details of the HH objects found in our recent study (MR04) of extended outflows from CTTSs.
Angular separation and position angle (P.A.) are with respect to the suggested source. 
The results of proper motion studies of these HH objects are given in \S\ref{sec-results4}. 
\newline $^a$ : Suggested source as given in MR04.
\newline $^b$ : While the HH828 knots are aligned with the extended CW Tau outflow we suggest in MR04 that, 
based on their morphology, they are not driven by CW Tau.
\newline $^c$ : HH\,834 is in the DO Tau/HV Tau C region however no candidate driving sources could be 
suggested in MR04.
}
\label{HHPositions}
\end{table}

\subsection{CW Tau}
\label{sec-CWTAU}

The HH\,220 bipolar ``micro\,-\,jet'' from CW Tau was first discovered by \cite{GomezdeCastro93}. \cite{Hirth94b} show this 
outflow to be 4\arcsec\ -- 6\arcsec\ on either side of CW Tau. 
In MR04 we presented a number of HH objects which appear to be an extension to this outflow. HH\,826\,A, HH\,826\,B and 
HH\,827 are to the south of CW Tau, with HH\,826\,C, HH\,828 and HH\,829 to the north (see Fig.\ \ref{cwtau_flow_pms}). In 
MR04 we assumed precession of the outflow over time, which gives the outflow an elongated inverted `S' shape i.e. that it  
consisted of HH\,826, HH\,827 and \,HH829. This suggested a total projected length for the outflow of $\sim$ 1 pc. 
The source of the HH\,828 trio of knots was unclear. Their alignment with the HH\,220 jet 
suggests that they are part of the CW Tau outflow, however the spatial spread of the knots could only occur if the outflow 
became much less collimated between HH\,826\,C and HH\,828. As the southern outflow from CW Tau remains collimated over a 
larger distance (to HH\,827) it is probable that HH\,828 is not driven by CW Tau.\\ 

Using our cross-correlation method we measure the proper motions of HH\,826 to HH\,829 in both [SII] and H$\alpha$. These 
objects are fainter in H$\alpha$, with the exception of HH\,827 which is much stronger, while HH\,826\,A and HH\,828 are not
seen at all in H$\alpha$ emission. The errors in alignment of the central frame (HH\,826, HH\,220\,NW and HH\,828) in [SII] 
and H$\alpha$ are quite high due to the low number of stars.\\

The P.A. of the blueshifted HH\,220 jet is at $\sim$ 144\degr\ with respect to CW Tau \citep{GomezdeCastro93,Dougados00} and 
we previously (MR04) measured the redshifted jet to be at $\sim$ 329\degr\ from the images of \cite{Dougados00}. 
From Table\ \ref{pms_cwtau} it can be seen that HH\,826\,A, which is quite close to the source, is moving in a direction of 
159\degr\ i.e. it is well aligned with the blueshifted HH\,220 jet. HH\,826\,B is moving at 156\degr\ in [SII] and 166\degr\ 
in H$\alpha$ and so is well aligned with both  HH\,826\,A and the blueshifted jet. Similarly HH\,826\,C, moving at 321\degr\
(in [SII]), is well aligned with the redshifted HH\,220 jet. This knot appears to be moving $\sim$ 65\degr\ further east (at
26\degr) in H$\alpha$.\\

\begin{table*}
\centering
\begin{tabular}{llcclcl}
\hline \hline
                &                &          & \multicolumn{2}{c}{[SII]}         & \multicolumn{2}{c}{H$\alpha$}      \\
HH Object       &Source $^a$     &Ang.      &Velocity      &Direction    &Velocity     &Direction                    \\ 
                &                &Sep.      &/kms$^{-1}$   &\ \ \ /\degr &/kms$^{-1}$  &\ \ \ /\degr                \\ \hline
HH\,826\,A      &CW Tau (B)      &0\farcm4  &132     &159 $\pm$ 24       &             &                              \\ 
HH\,826\,B      &CW Tau (B)      &1\farcm2  &207     &156 $\pm$ 15       &282          &166 $\pm$ 11                  \\ 
HH\,827\,A      &CW Tau (B)      &6\farcm1  &        &                   &347          &180 $\pm$ (2 - 5)             \\
HH\,827\,B      &CW Tau (B)      &6\farcm9  &199     &198 $\pm$ (3 - 8)  &151          &188 $\pm$ (4 - 11)            \\ 
HH\,220\,NW     &CW Tau (R)      &0\farcm6  &323     &287 $\pm$ 10       &             &                     \\ 
HH\,826\,C      &CW Tau (R)      &0\farcm8  &261     &321 $\pm$ 12       &111          &26 $\pm$ 29                   \\
HH\,828 M       &                &          &118     &144 $\pm$ 27       &             &                              \\ 
HH\,828 W       &                &          &173     &208 $\pm$ 18       &             &                               \\ 
HH\,829\,A      &                &          &116     &108 $\pm$ (5 - 14) &             &                               \\    
HH\,829\,B      &                &          &113     &163 $\pm$ (5 - 14) &             &                            \\ \hline
\end{tabular}
\caption {Tangential velocity and direction of motion of HH\,220, HH\,826, HH\,827, HH\,828 and HH\,829 in the vicinity of 
CW Tau assuming a distance of 140 pc to the source \citep{Elias78,Wichmann98}. The associated errors for HH\,826, 
HH\,220\,NW and HH\,828 are $\pm$ $\sim$ 55 kms$^{-1}$. For HH\,827 and HH\,829 the errors are $\pm$ (10 -- 28) kms$^{-1}$
(see \S\ref{sec-propermotions}). All errors are 1$\sigma$. For Tables\ \ref{pms_cwtau} -- \ref{pms_rwaur} the discrepancies
between the [SII] and H$\alpha$ measurements are discussed in the text and in \S\ref{sec-diff_velocities}.
\newline $^a$ : For the HH objects that are driven by CW Tau, B and R denotes whether the object is aligned with the blue or 
redshifted lobe.} 
\label{pms_cwtau}
\end{table*}

There are two optically visible parts to the extended redshifted HH\,220 jet. First, it extends to 9\arcsec\ from the source,
there is then a gap before the second part of the jet is seen at 20\arcsec\ to 37\arcsec\ from the source, ending in 
HH\,826\,C (MR04). The 9\arcsec\ long segment close to CW Tau was too faint to measure, however the longer HH\,220\,NW 
(northwest) segment was measured in [SII] and has a direction of 287\degr. This direction 
is roughly aligned with HH\,826\,C (at 321\degr) and with the P.A. of the redshifted ``micro\,-\,jet'' closer to the source 
(329\degr).\\

For the purpose of more accurate proper motions, we have divided HH\,827 into Knot B (the brightest and largest part of 
HH\,827) and Knot A (the closest bright knot to CW Tau) -- see Fig.\ \ref{cwtau_flow_pms}. HH\,827 is much stronger in 
H$\alpha$, and these proper motion studies show it to be moving at 184\degr\ (this is an average for HH\,827\,A \& B in 
H$\alpha$). Although HH\,827\,B is much smaller and fainter in [SII], its direction of motion is still in good agreement 
with the H$\alpha$ values at 
198\degr. Knot A is too faint to measure in [SII]. Even taking errors into account, these directions rule out 
IRAS\,04113+2758 (see Fig.\ \ref{cwtau_flow_pms}) as a possible source of HH\,827, as suggested in MR04. As we do not know 
if and where this outflow starts to bend beyond HH\,826\,B it is difficult to know if this object is part of the CW Tau 
outflow. It was suggested in MR04 that CW Tau could be driving this object if the outflow is precessing and these results 
are consistent with this scenario.\\

The velocity of HH\,827\,B is 199 kms$^{-1}$ in [SII] and 151 kms$^{-1}$ in H$\alpha$. These values are just outside the 
associated errors.  
There appears to be a very large decrease in velocity from 347 kms$^{-1}$ for Knot A to 151 kms$^{-1}$ for Knot B.\\ 

The HH\,828 knots to the north of CW Tau are quite faint and diffuse and their proper motions are not easily measured. 
The western and middle knots are moving at approximately 208\degr\ and 144\degr\ respectively while it was not possible to 
measure proper motions for the eastern knot.These directions rule out the possibility of them being driven by CW Tau, 
IRAS\,04111+2804 or IRAS\,04108+2803 (see Fig.\ \ref{cwtau_flow_pms}) as suggested in MR04. There are two candidate driving 
sources to the north of HH\,828: FN Tau (a CTTS) and IRAS\,04110+2820.
HH\,828 is 13\arcmin\ at a P.A. of 184\degr\ from FN Tau, and is 13\farcm1 at a P.A. of 181\degr\ from IRAS\,04110+2820. 
Both of these sources are just beyond the northern field of view of Fig.\ \ref{cwtau_flow_pms}. The proper motion directions
of the HH\,828 knots suggest that they may by driven by either of these sources. However HH\,828 is at a projected 
distance of $\sim$0.5pc ($\sim$13$'$) and at that distance from these possible sources we would expect that only the 
largest, most chaotic shocks survive.\\ 

Turning to HH\,829, knot C is too faint to measure in both [SII] and H$\alpha$ and knots A and B were measured in [SII] 
only. Knot A is moving in a direction of 108\degr\ and B is at 163\degr, ruling out CW Tau as their driving source. 
It can be seen from Fig.\ \ref{cwtau_flow_pms} that HH\,829\,B has a similar direction of motion to that of the HH\,828 
knots, suggesting a common driving source. HH\,829\,B is 1\farcm8 from FN Tau and 1\arcmin\ from IRAS\,04110+2820. 
It is possible that HH\,829\,B is part of an outflow driven by FN Tau or IRAS\,04110+2820 (as mentioned earlier in respect 
to HH\,828). There is no evidence of 
jets emanating from either source.\\

\begin{figure*}[!htp]
\resizebox{\hsize}{!}{\includegraphics{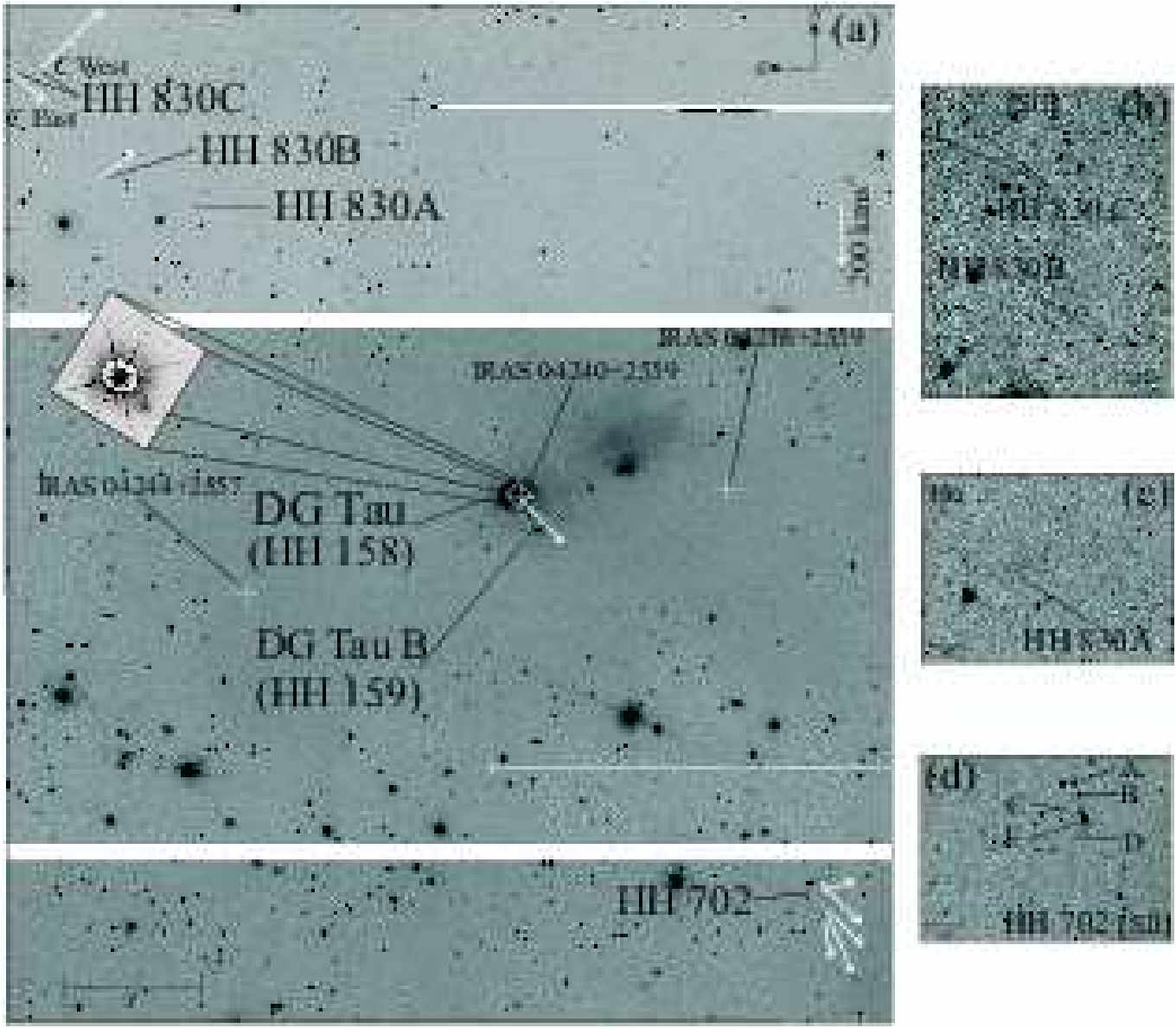}}
\caption{{\bf a)} Mosaic image showing proper motions of HH\,702 and HH\,830 in the vicinity of DG Tau in [SII]. The 
direction of motion of each HH object is represented by white arrows. The relative length of these arrows denotes the 
relative velocity of the object. Our study shows that HH\,702 is likely to be driven by DG Tau, but that HH\,830 is not. 
HH\,830 C (West) and B appear to be driven by a source to the east however we are unable to suggest any. As HH\,830\,C is on
the edge of the mosaic, its proper motion is possibly less precise. A HST image of HH\,158 is included here as an inset. 
{\bf b)} HH\,830\,B and C in [SII]. The error in velocity for these objects are high ($\pm$ 60 km$^{-1}$ in [SII] and 
$\pm$ 40 kms$^{-1}$ in H$\alpha$, see \S\ref{sec-propermotions}) as they are very faint. {\bf c)} HH\,830\,A is 
stronger in H$\alpha$ than in [SII], however it is still too faint to do proper motion studies on. {\bf d)} Knots A -- E of 
HH\,702 (See MR04).}
\label{dgtau_flow_pms}
\end{figure*}

\begin{table*}
\centering
\begin{tabular}{llcclcl}
\hline \hline
                &        &          & \multicolumn{2}{c}{[SII]}           & \multicolumn{2}{c}{H$\alpha$}               \\
HH object       &Source  &Ang.      &Velocity     &Direction              &Velocity    &Direction                       \\ 
                &        &Sep.      &/kms$^{-1}$  &\ \ \ /\degr           &/kms$^{-1}$ &\ \ \ /\degr                \\ \hline
HH\,158\,C $^a$ &DG Tau  &12\arcsec &197          &223                    &            &                                 \\ 
HH\,702\,A      &DG Tau  &10\farcm6 &203          &225 $\pm$ (3 - 8)      &129         &194 $\pm$ (5 - 12)               \\ 
HH\,702\,B      &DG Tau  &10\farcm9 &97           &285 $\pm$ (3 - 8)  \\ 
HH\,702\,C      &DG Tau  &11\farcm1 &204          &224 $\pm$ (6 - 17)     &149         &128 $\pm$ (4 - 11)               \\ 
HH\,702\,D      &DG Tau  &11\farcm4 &219          &205 $\pm$ (3 - 8)      &190         &269 $\pm$ (3 - 9)                \\ 
HH\,702\,E      &DG Tau  &11\farcm4 &125          &191 $\pm$ (3 -7)       &186         &117 $\pm$ (3 - 9)                \\
HH\,830\,C West &        &          &298          &314 $\pm$ 28           &            &                                 \\
HH\,830\,C East &        &          &161          &226 $\pm$ 12           &348         &314 $\pm$ 7                      \\
HH\,830\,B      &        &          &142          &313 $\pm$ 22           &327         &233 $\pm$ 7                \\ \hline
\end{tabular}
\caption
{Tangential velocity and direction of motion of HH\,158, HH\,702 and HH\,830 in the vicinity of DG Tau assuming a distance 
of 140 pc \citep{Elias78,Wichmann98}. The associated errors for HH\,830 are $\pm$ 60 kms$^{-1}$ in [SII] and $\pm$ 40 
kms$^{-1}$ in H$\alpha$ images. For HH\,702 the errors are $\pm$ (10 -- 28) kms$^{-1}$ (see \S\ref{sec-propermotions}). 
All errors are 1$\sigma$.
\newline $^a$ : Proper motions could not be obtained by us for HH\,158 (see text). Angular separation and velocity values 
are taken from \cite{Eisloffel98} and the direction of motion of HH\,158 is taken from \cite{Bacciotti02}.}
\label{pms_dgtau}
\end{table*}

It can be seen from the results in Table\ \ref{pms_cwtau} that the velocity measurements can be quite different for 
individual knots in H$\alpha$ and [SII]. However the difference in [SII] and H$\alpha$ proper motion vectors for many of 
the objects are within errors. For example, HH\,826\,B has a velocity and position angle difference of 75 kms$^{-1}$ and 
10\degr.  The errors for this frame are quite large as mentioned earlier (due to the small number of stars) i.e. $\pm$ 55 
kms$^{-1}$ for both the [SII] and H$\alpha$ measurement while the sum of the errors in position angle is 26\degr.
For HH\,827\,B, the difference in velocity measurements is just outside the associated errors while that of the position 
angle is within the errors. HH\,826\,A, however, has substantial differences between its [SII] and H$\alpha$ measurement. 
This point is further discussed in \S\ref{sec-diff_velocities}.
For the rest of the HH objects listed in Table\ \ref{pms_cwtau} there are only values for either [SII] or H$\alpha$.\\

While this proper motion study confirms that HH\,826 and HH\,827 are likely driven by CW Tau, it shows that HH\,828 and 
HH\,829 are not. This reduces the known projected length of the outflow from 0.98pc (24\arcmin), as suggested in MR04, to 
0.32pc (7\farcm7), which is still much greater than the $\sim$ 0.008pc ($\sim$ 12$''$) long ``micro\,-\,jet'' that was 
previously seen.\\

\subsection{DG Tau}
\label{sec-DGTAU}

HH\,158 is a ``micro\,-\,jet'' close to DG Tau that was originally discovered by \cite{Mundt83} and extends for $\sim$ 
16$''$ \citep{Eisloffel98} at a P.A. of 223\degr\ \citep{Bacciotti02}. In MR04 we suggested that the DG Tau outflow is much 
longer and may also consist of two other HH complexes which are approximately aligned with HH\,158 -- HH\,702 which 
was also independently discovered by \cite{Sun03} and HH\,830 (see Fig.\ \ref{dgtau_flow_pms}). If this were the case the
DG Tau outflow would actually extend for at least 27\arcmin\ ($\sim$1.1 pc). Here we present proper motion studies 
confirming that HH\,702 is indeed driven by DG Tau, however our studies of HH\,830 suggest it is not.\\

Table\ \ref{pms_dgtau} shows the velocity and direction of motion of knots A -- E in HH\,702. 
In [SII] emission knots A, C, D and E are moving at between $\sim$ 191\degr\ and $\sim$ 225\degr. These knots are well 
aligned with HH\,158 which has as a P.A. of 223\degr\ with respect to DG Tau (see Fig.\ \ref{dgtau_flow_pms}). Knot B 
appears to be moving at 285\degr\ and so is not aligned with the other knots in HH\,702. Knots A, C and D have comparable 
velocities between $\sim$ 203 kms$^{-1}$ and $\sim$ 219 kms$^{-1}$ while knots B and E have velocities of only approximately
half of this, at 97 kms$^{-1}$ and 125 kms$^{-1}$ respectively. It can be clearly seen from Fig.\ \ref{dgtau_flow_pms} that 
knots A to E (excluding B) are moving in a direction that would be expected if DG Tau is the driving source. Hence, we 
can rule out the possibility that knot B has a different source.\\ 

The direction of motion found in H$\alpha$ for the HH\,702 knots range from 117\degr\ for knot E to 269\degr\ for 
knot D. Knot A is moving within $\sim$ 30\degr\ of the direction found in [SII], which is outside the associated errors for 
for the two measurements, while knots C, D and E are significantly different. 
Knots A and C have comparable velocities of $\sim$ 129 kms$^{-1}$ and 149 kms$^{-1}$ in H$\alpha$ while knots D and E are 
also comparable at $\sim$ 186 kms$^{-1}$ and 190 kms$^{-1}$. Only knot D has comparable velocity in both H$\alpha$ and [SII] 
of 190 kms$^{-1}$ and 215 kms$^{-1}$. Knot B has a much slower velocity of 45 kms$^{-1}$ in H$\alpha$.\\ 

For the HH\,702 knots A, C, D and E we have obtained proper motions in both H$\alpha$ and [SII] emission lines. HH\,702\,B 
is too faint to get proper motions in H$\alpha$. Calculating the 
{\it average} velocities and direction of motions for these four knots yields comparable values of 188 kms$^{-1}$ at 
211\degr\ in [SII] and 164 kms$^{-1}$ at 177\degr\ in H$\alpha$. The direction of motion for HH\,702 
knots A -- E is in very good agreement with the P.A. of the HH\,158 jet at 223\degr.\\ 

HH\,830 to the northeast of DG Tau consists of knots A, B and C, with knot C containing two separate emission regions (C 
West and C East). Table\ \ref{pms_dgtau} shows the results of the proper motion study of HH\,830\,B and C only as knot A is 
too faint to measure. Fig.\ \ref{dgtau_flow_pms} shows that knots B and C West are moving in parallel. Knot C East is moving
orthogonally to these however this knot is at the eastern edge of the field of view of the WFC CCD mosaic which makes it 
more difficult to determine accurate proper motions. Also, as mentioned in \S\ref{sec-propermotions}, the errors in the 
proper motions of HH\,830 in both [SII] and H$\alpha$ are quite high. These 
direction of motions prove that these knots are not driven by DG Tau. 
Therefore the projected length of the DG Tau outflow must be revised downwards to 0.5 pc (12\farcm33).\\  
 
We were unable to obtain proper motions for HH\,158 from our images due to the significant amount of nebulosity 
surrounding this object. \cite{Eisloffel98} determined the velocity of Knot C of HH\,158 as 197 kms$^{-1}$. 
The results from our proper motion study show that the velocity of HH\,702 is comparable to that of HH\,158 (both in [SII]),
 despite the 7\farcm7 (0.3 pc projected) gap between these two objects. This point is discussed further in 
\S\ref{sec-velofextended}.

\subsection{The DO Tau and HV Tau C region}
\label{sec-DOHVTAU}

\begin{figure*}[!htp]   
\resizebox{\hsize}{!}{\includegraphics{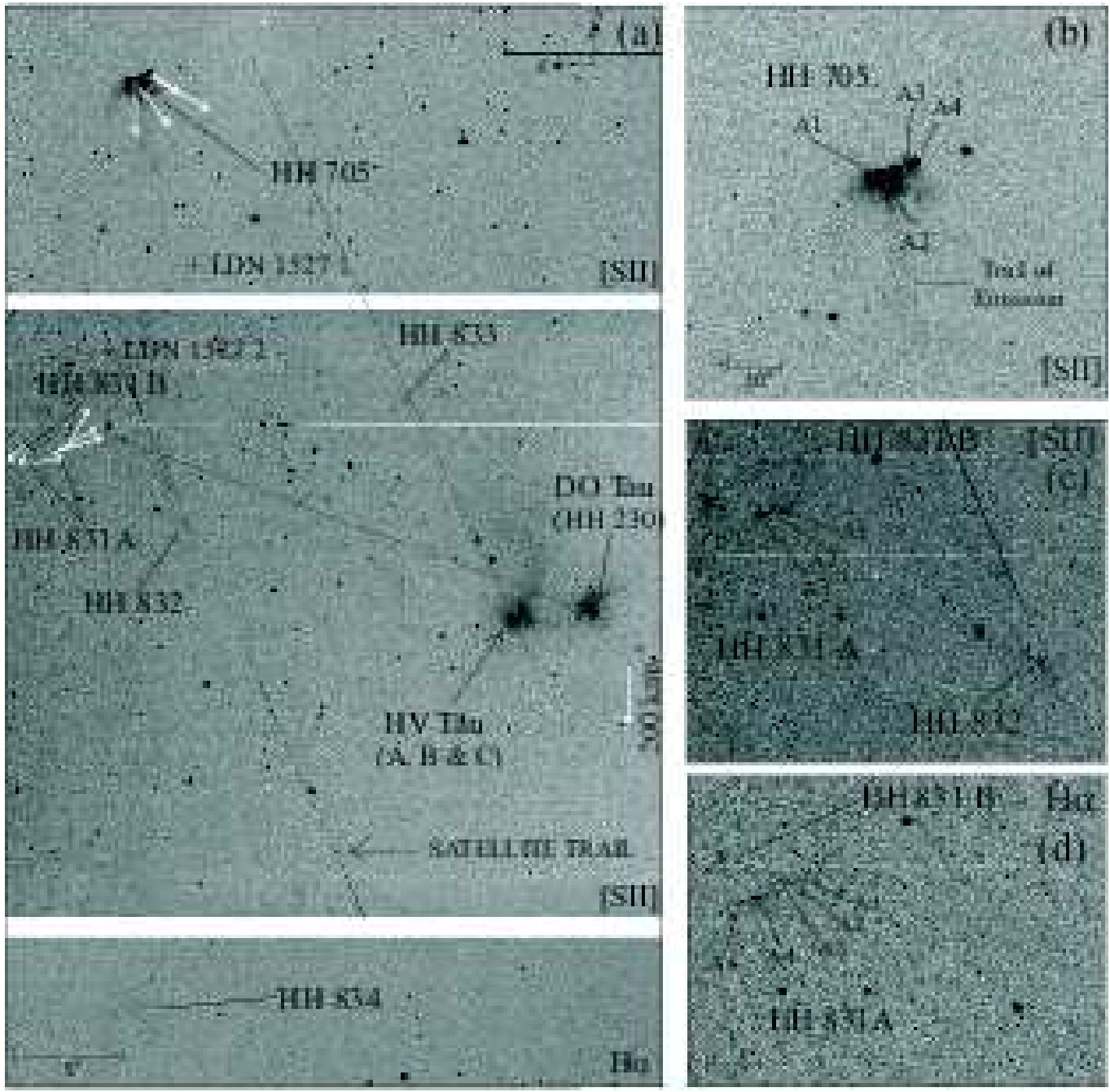}}            
\caption{{\bf a)} Mosaic image showing proper motion vectors for the HH objects in the vicinity of DO Tau and HV Tau C. 
The direction of motion of each HH object is represented by white arrows. The relative length of these arrows denotes the
relative velocity of the object. These vectors show, surprisingly, that neither of these sources are driving HH\,831 and 
HH\,705 as was suggested in MR04. No proper motions could be detected for the fainter objects HH\,832, HH\,833 and HH\,834, 
however it is still possible that HH\,832 may be driven by DO Tau based on its rough alignment with DO Tau's jet. 
The dashed line from DO Tau is at 70\degr\ which is the P.A. of HH\,230 while the dashed line from HV Tau C is at 25\degr, 
marking the P.A. of its ``micro\,-\,jet''. Candidate driving sources for HH\,831, HH\,833 and HH\,705 are discussed in the 
text. Images {\bf b -- d} show the individual knots in HH\,705 and HH\,831 for which the proper motion vectors were
calculated.}                  
\label{dotau_flow_pms}
\end{figure*}

\begin{table*}
\centering
\begin{tabular}{lccclcl}
\hline \hline
             &       &     & \multicolumn{2}{c}{[SII]}       & \multicolumn{2}{c}{H$\alpha$}                            \\
HH Object    &Source &Ang. &Velocity     &Direction          &Velocity       &Direction                                 \\ 
             &       &Sep. &/kms$^{-1}$  &\ \ \ /\degr       &/kms$^{-1}$    &\ \ \ /\degr                         \\ \hline
HH\,831\,A1  &$^a$   &     &184          &331 $\pm$ (3 - 9)  &               &                                          \\ 
HH\,831\,A2  &$^a$   &     &166          &300 $\pm$ (4 - 10) &               &                                          \\ 
HH\,831\,A3  &$^a$   &     &196          &288 $\pm$ (3 - 8)  &               &                                          \\ 
HH\,831\,A4  &       &     &             &                   &141            &32 $\pm$ (4 - 11)                         \\
HH\,831\,A5  &$^a$   &     &112          &284 $\pm$ (5 - 14) &221            &182 $\pm$ (3 - 7)                         \\
HH\,831\,B1  &       &     &71           &158 $\pm$ (8 - 22) &               &                                          \\ 
HH\,831\,B2  &       &     &79           &134 $\pm$ (8 - 20) &               &                                          \\ 
HH\,705\,A1  &       &     &187          &180 $\pm$ (3 - 9)  &330            &186 $\pm$ (2 - 5)                         \\ 
HH\,705\,A2  &       &     &147          &218 $\pm$ (4 - 11) &231            &186 $\pm$ (3 - 7)                         \\ 
HH\,705\,A3  &       &     &244          &241 $\pm$ (3 - 7)  &194            &131 $\pm$ (3 - 8)                         \\ 
HH\,705\,A4  &       &     &99           &232 $\pm$ (6 - 16) &211            &227 $\pm$ (3 - 8)                    \\ \hline
\end{tabular}
\caption
{Tangential velocity and direction of motion of HH\,831 and HH\,705 assuming a distance of 140 pc \citep{Elias78,Wichmann98}
to the Taurus Auriga cloud. The associated errors for HH\,831 and HH\,705 are $\pm$ (10 -- 28) kms$^{-1}$ (see 
\S\ref{sec-propermotions}). All errors are 1$\sigma$.
\newline $^a$ : The driving source is currently unknown however three IRAS sources are suggested in the text as possible 
sources for HH\,831\,A1 -- A3 and A5.}
\label{pms_dotau_hvtau}
\end{table*}

The 2\arcsec\ -- 4\arcsec\ long bipolar HH\,230 ``micro\,-\,jet'' at a P.A. of $\sim$ 70\degr\ with respect to DO Tau was 
first observed by \cite{Hirth94}. Two newly discovered HH objects, HH\,831 and HH\,832 (see MR04), to the east of DO Tau 
appear to be well aligned with HH\,230 (see Fig.\ \ref{dotau_flow_pms}), suggesting that they are part of this outflow. 
However this proper motion study shows the HH\,831 knots (A and B) to be moving in a completely different direction than 
that expected if driven by DO Tau.\\ 

HH\,831\,A knots A1, A2, A3 and A5 are moving in similar directions between 284\degr\ and 331\degr\ 
(Table\ \ref{pms_dotau_hvtau}) in [SII]. Knot A4 is too faint to measure in [SII]. B1 and B2 (only measured in [SII]) 
however are moving in a different direction at 158\degr\ and 134\degr\ respectively. Only knots A4 and A5 could be measured 
in H$\alpha$ however the directions of knot A4 at 32\degr\ is not aligned with any of the directions seen in [SII] 
and knot A5 is moving at 182\degr\ in H$\alpha$, which is not aligned with its [SII] results.  
As the measurable motions in H$\alpha$ very different from [SII] and the rest of HH\,831\,A and HH\,831\,B are too faint 
to be detected in H$\alpha$ it would appear that the H$\alpha$ results are not reliable.\\ 

The velocities of HH\,831\,A1, A2 and A3 (in [SII]) are comparable and range from 166 kms$^{-1}$ to 196 kms$^{-1}$ with knot
A5 slower at 112 kms$^{-1}$. Knots B1 and B2 have comparable velocities of 71 kms$^{-1}$ and 79 kms$^{-1}$ respectively.\\

From these results it is obvious that HH\,831 is not driven by DO Tau, despite the fact that these knots appear to be very 
well aligned with the known P.A. of HH\,230 (Fig.\ \ref{dotau_flow_pms}). These results show a number of different 
directions of motion within HH\,831 -- knots A1, A2, A3 and A5 (in [SII]) are moving west/northwest, while knots B1 and B2 
are moving to the southeast. Knot A4 (in H$\alpha$) appears to be separate to all of these and is moving to the northeast. 
Possible driving sources for HH\,831\,A based on these proper motion results will be discussed at the end of this section.
It is possible that either of the two nearby LDN\,1527 radio sources \citep{Anglada92} could be driving HH\,831\,B1 and B2 
(see Fig.\ \ref{dotau_flow_pms}). HH\,832, which is seen in [SII] only, is too faint to measure its proper motion.\\

The proper motions of HH\,831\,A (between 284\degr\ and 331\degr) may suggest a common source to its east. There are six 
known candidates. The distance and P.A. of HH\,831\,A relative to each
of these is given below followed by a discussion on the most likely source:\\ 
1. ITG 14, a YSO \citep{Itoh99} (6\arcmin\ at a P.A. of 275\degr)\\
2. IRAS 04365+2605 (6\farcm5 at a P.A. of 291\degr)\\ 
3. IRAS 04371+2559 (16\farcm8 at a P.A. of 299\degr)\\ 
4. IRAS 04370+2559 (15\farcm7 at a P.A. of 302\degr)\\ 
5. IRAS 04368+2557 (13\farcm8 at a P.A. of 312\degr)\\ 
6. IRAS 04366+2556 (13\farcm5 at a P.A. of 325\degr)\\ 
The most likely source is IRAS\,04368+2557, which has previously been associated with HH\,192\,A, B and C 
\citep{Eiroa94,Gomez97}. From the images of \cite{Gomez97} HH\,192\,A and B are aligned with the blue lobe of a molecular 
outflow and extend for $\sim$ 2\arcmin\ from the IRAS source at a P.A. of $\sim$ 80\degr. HH\,192\,C is aligned with the 
redshifted lobe of the molecular outflow and is $\sim$ 4\farcm2 from IRAS\,04368+2557 at a P.A. of 277\degr. If the outflow 
is precessing HH\,831\,A could be part of it.\\

A 1\farcs5 long bipolar ``micro\,-\,jet'' was recently discovered from HV Tau C by \cite{Stapelfeldt03}, which we estimate
to be at a P.A. of 25\degr\ (northern jet) from their images (MR04). HH\,833, to the northeast of HV Tau C, is also at a
P.A. of 25\degr\ suggesting that it may also be driven by this source (see Fig.\ \ref{dotau_flow_pms}). Further out in this 
direction is HH\,705 (independently discovered by Sun et al., 2003) which may also be part of this outflow. Due to  
morphological alignment we suggested in MR04 that LDN 1527 1 and LDN 1527 2 which lie to the south of HH\,705 are other 
possible driving sources of HH\,705 (MR04; Sun et al., 2003) but we suggested that it was unlikely that LDN 1527 1 could 
generate such a large object so nearby.\\
 
Proper motion studies of the HH\,833 and HH\,705 objects did not reveal anything conclusive about their sources. HH\,833 was 
too faint to measure, so we still suggest that it is driven by HV Tau C due to its exact alignment with the HV Tau C jet 
(however see below for an alternative driving source). Fig.\ \ref{dotau_flow_pms} shows the direction of motion of the 
HH\,705 knots which are detailed in Table\ \ref{pms_dotau_hvtau}. All knots in HH\,705 (A1 -- A4) are moving to the 
south/southwest in [SII]. The H$\alpha$ results show a similar direction of motion however A3 is moving in a more 
southwesterly direction (at 131\degr). The southward motions of HH\,705 conclusively rule out HV Tau C, LDN\,1527\,1 or 
LDN\,1527\,2 as its source (see MR04 and Fig.\ \ref{dotau_flow_pms}). 
There is an IRAS source, IRAS\,04358+2618, $\sim$ 5\arcmin\ (1.5 pc) to the north of HH\,705 which remains the only known 
possible driving source for this object. HH\,705 is at a P.A. of 213\degr\ with respect to this source which is well aligned 
with the proper motion of the HH\,705 knots which are between 180\degr\ and 241\degr. HH\,833 is at a P.A. of 221\degr\ with 
respect to HH\,705 showing that IRAS\,04358+2618 could be driving a $\sim$12\farcm7 long outflow consisting of HH\,705 and 
HH\,833. \\

The HH\,705 knots A1 -- A4 have been measured in both [SII] and H$\alpha$. While the individual knots show different 
velocities and direction of motion, if we get the average for both of these emission lines the proper motions are presumably 
more reliable: 169 kms$^{-1}$ at 218\degr\ in [SII] and 242 kms$^{-1}$ at 183\degr\ in H$\alpha$.\\ 

Our proper motion studies have shown that HH\,705 and HH\,831 are not driven by DO Tau or HV Tau C. In MR04 
we suggested that these CTTSs were driving parsec\,-\,scale outflows however our proper motion study shows that this is not 
the case.\\

\subsection{RW Aur}

\begin{figure}
\resizebox{\hsize}{!}{\includegraphics{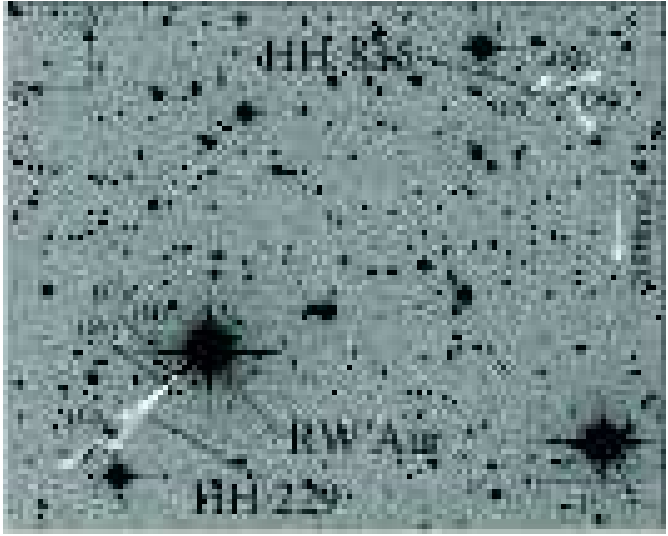}}       
\caption{Proper motion vectors for HH\,835 and HH\,229 in the RW Aur outflow in H$\alpha$. The direction of motion 
of each HH object is represented by white arrows. The relative length of these arrows denotes the relative velocity of the 
object. The direction of motion of the eastern and middle knots of HH\,835 show RW Aur to be driving this object. Proper 
motions were also measured for four knots in the blueshifted HH\,229 outflow.}
\label{rwaur_flow_pms}
\end{figure}

The HH\,229 jet from RW Aur was first noted by \cite{Hamann94} and \cite{Hirth94}. The outflow is at a P.A. of 130\degr\ 
\citep{Dougados00,Mundt98,Hirth97} with respect to RW Aur and is at least 106\arcsec\ in length while the redshifted outflow
is at least 50\arcsec\ long \citep{Mundt98}. HH\,835 is 5\farcm37 from RW Aur at a P.A. of 310\degr\ which is exactly 
aligned with the redshifted HH\,229 jet (see Fig.\ \ref{rwaur_flow_pms}). HH\,835 is much stronger in H$\alpha$ 
than [SII] and proper motions were only determined for H$\alpha$. In MR04 we suggested that it is a bow shock with 
only the northern wing visible here. If HH\,835 is part of this outflow then the total projected length of this outflow from 
the blueshifted HH\,229 jet to HH\,335 is 0.29 pc (7\arcmin).\\

\begin{table}
\centering
\begin{tabular}{llcclcl}
\hline \hline
                &              &             & \multicolumn{2}{c}{H$\alpha$}                                       \\
HH Object       &Source $^a$   &Ang.         &Velocity          &Direction                                         \\ 
                &              &Sep.         &/kms$^{-1}$      &\ \ \ /\degr                                       \\ \hline
HH\,835\,E      &RW Aur (R)    &4\farcm5     &92                 &302 $\pm$ (6 - 17)                               \\
HH\,835\,M      &RW Aur (R)    &4\farcm6     &208                &288 $\pm$ (3 - 8)                                \\
HH\,835\,W      &RW Aur (R)    &4\farcm7     &162                &216 $\pm$ (4 - 10)                               \\
HH\,229\,B      &RW Aur (B)    &0\farcm4     &223                &132 $\pm$ (3 - 7)                                \\
HH\,229\,C      &RW Aur (B)    &0\farcm5     &307                &139 $\pm$ (2 - 5)                                \\
HH\,229\,D      &RW Aur (B)    &0\farcm6     &238                &127 $\pm$ (3 - 7)                                \\
HH\,229\,G      &RW Aur (B)    &1\farcm6     &235                &126 $\pm$ (3 - 7)                                \\ \hline
\end{tabular}
\caption
{Tangential velocity and direction of motion of HH\,835 and HH\,229 in the RW Aur outflow assuming a distance of 140 pc to 
the source \citep{Elias78,Wichmann98}. HH\,835 is very weak in [SII] so proper motion studies could only be done with the 
H$\alpha$ images. The nomenclature for the knots in HH\,229 are taken from \cite{Eisloffel98}. The associated errors for 
HH\,229 and HH\,835 are $\pm$ (10 -- 28) kms$^{-1}$ (see \S\ref{sec-propermotions}). All errors are 1$\sigma$.
\newline $^a$ : B and R denotes whether the object is aligned with the blue or redshifted jet.}                
\label{pms_rwaur}
\end{table}

HH\,835 appears to consist of a number of discrete objects which we refer to here as the west (W), middle (M) and east (E) 
knots in Fig.\ \ref{rwaur_flow_pms} and Table\ \ref{pms_rwaur}. Of these, the eastern knot appears to be the brightest. 
Proper motion studies show this knot to be moving in a direction of 302\degr\ (Table\ \ref{pms_rwaur}) which is well 
aligned with the P.A. of the redshifted HH\,229 jet. The motion of the middle knot is also closely aligned at 
288\degr. The western knot however is at 216\degr\ and so appears to be moving in a different direction. Overall, these 
proper motion vectors confirm the suggestion that this object is part of the RW Aur outflow. In MR04 we suggested that 
HH\,835 may be one side of a bow shock, with the other side being optically obscured. In this case, knot W would be at the 
head of the bow shock with knots M and E further back along one side of the shock. The velocities of knots W and M are 
162 kms$^{-1}$ and 208 kms$^{-1}$ while knot E is has a much lower velocity of 92 kms$^{-1}$. These values are consistent
with a slowing down in velocity of knots along the edge of the bow shock away from the head.\\

Proper motions were also obtained for some of the brighter knots in the HH\,229 blueshifted outflow. 
Knots B, C, D and G (using the nomenclature of Eisl{\" o}ffel \& Mundt, 1998) were 
strong enough in H$\alpha$ to determine their tangential velocity and direction -- see Table\ \ref{pms_rwaur}. 
Knots B, D and G can be seen from Fig.\ \ref{rwaur_flow_pms} to be moving along the optically visible jet direction at 
132\degr, 127\degr\  and 126\degr\ respectively, while knot C is slightly off this direction at 139\degr. The velocities of 
knots B, D and G are comparable, ranging from 223 kms$^{-1}$ to 238 kms$^{-1}$ while knot C is faster at 307 kms$^{-1}$. 
The redshifted counterflow was too faint to measure. 
The known angular extent of the RW Aur outflow remains at 7\arcmin\ (0.29pc).

\section{Discussion}
\label{sec-discussion}
\subsection{Determining Outflow Sources}
The proper motion studies undertaken here have confirmed the sources suggested in MR04 for a number of HH objects e.g. 
HH\,826 and HH\,827 (CW Tau); HH\,702 (DG Tau) and HH\,835 (RW Aur). However they have also refuted the suggested sources 
for others -- HH\,828, HH\,829, HH\,830, HH\,831 and HH\,705. This was surprising in some cases as the objects 
were well aligned with known outflows -- this is most noticeable in the case of DO Tau where HH\,831 appeared to be an 
extension of the HH\,230 outflow based on their close alignment. In these cases, we have tried to suggest candidate sources 
based on their newly discovered direction of motion.\\ 

These proper motion studies highlight the need for circumspection in using the apparent alignment of HH objects to derive 
their driving source. Nevertheless, in the {\em absence} of proper motion studies apparent outflow alignment is {\em still 
the best means} of finding potential driving sources. Proper motion studies can then conclusively confirm or refute these 
sources at a later date. As the CTTSs examined in this paper are relatively close by at 140pc, a time difference between 
observations of just a few years is sufficient to accurately measure their motion. 

\subsection{Parsec-Scale Outflows from CTTSs}
In MR04 we suggested that five CTTS sources (CW Tau, DG Tau, DO Tau, HV Tau C and RW Aur) drive large\,-\,scale outflows, of
the order of 0.5pc -- 1pc, based on distant HH objects that appeared morphologically and/or by position to be part of the 
outflow.  However in light of the present study	we now know that only three of these stars definitely do: CW Tau drives a 
7\farcm7 (0.32pc) outflow, DG Tau drives a 12\farcm3 (0.5pc) outflow while the known angular extent of the RW Aur outflow is
still 7\arcmin\ (0.29pc). While the length of these outflows is at the lower limit of those suggested in MR04, they are 
still much larger than the ``micro\,-\,jets'' they were previously known to drive. HV Tau may still be associated with 
HH\,833 which was too faint to measure its proper motion.\\
 
The CW Tau and DG Tau outflows appear very much one-sided in that the blue side is extended while the red side appears very, 
very short. It is not uncommon for the red side of an outflow to be hidden from view as it recedes from us into the cloud,  
so it is very likely that the CW Tau and DG Tau outflows are more extended in the red direction than our images show. 

\subsection{[SII] verses H$\alpha$ Velocities}
\label{sec-diff_velocities}
Proper motions were determined in both [SII] and H$\alpha$ for the majority of HH objects here, the main exception 
being HH\,835 and HH\,229 where the individual knots were too weak in [SII] images to measure. There are few proper motion 
studies which calculate proper motions for both [SII] and H$\alpha$ simultaneously and of these some find velocity 
measurements which are very similar e.g. HH\,34 \citep{Reipurth02} while others find large variations for some knots 
e.g. HH\,157 driven by FS Tau B \citep{Eisloffel98}. 
Our results show varying degrees of agreement between H$\alpha$ and [SII]; some knots have comparable 
velocities/directions of motion in these two emission lines, while others are quite different. However some of these 
differences can be understood in terms of errors in their measurements. 
Also, \cite{Hartigan01} note that 
brightness changes between two different epochs will affect the H$\alpha$ image more than the [SII] as H$\alpha$ 
responds immediately to changes in the preshock density while [SII] averages this variability over the $\sim$ 30 year 
cooling time. The different timescales over which cooling acts may explain some of the  discrepancies in the results 
presented here.

\subsection{Velocity of Distant HH Objects}
\label{sec-velofextended}
In the introduction, we mentioned that the tangential velocity of objects at relatively large distances from their sources 
is currently not well known. In MR04 the dynamical timescales of these outflows were calculated using an assumed tangential 
velocity of 50 kms$^{-1}$, comparable to the lowest velocity necessary to collisionally excite the shocks observed. This 
assumed velocity yielded dynamical ages of 0.6 $\times$ 10$^4$ -- 2.1 $\times$ 10$^4$ years. However from examining 
Tables\ \ref{pms_cwtau},\ \ref{pms_dgtau} and\ \ref{pms_rwaur} we see that the velocity of the more distant objects is much 
higher than the 50 kms$^{-1}$ lower limit. HH\,827 (driven by CW Tau) has a value of 249 kms$^{-1}$ (average of Knots A and 
B) in H$\alpha$ and 199 kms$^{-1}$ (Knot B) in [SII]. Knots A, C, D and E of HH\,702 (driven by DG Tau) have an
average velocity of 188 kms$^{-1}$ in [SII] and of 164 kms$^{-1}$ in H$\alpha$. Finally HH\,835 (driven by RW Aur) has 
an average velocity of 154 kms$^{-1}$. Examination of these values would suggest that 200 kms$^{-1}$ is more typical for 
relatively distant HH objects driven by CTTSs.\\

This study has also shown that the lengths of these outflows are shorter than previously suggested, as some of the HH 
objects that appeared to be associated with the CTTSs are shown here to be unrelated. Re-calculating dynamical timescales to
take account of these results reduces them to 1.4 $\times$ 10$^3$ yr for CW Tau, 2.3 $\times$ 10$^3$ yr for DG Tau and 0.9 
$\times$ 10$^3$ yr for RW Aur. However we stress that these are much lower than the actual outflow timescales (McGroarty, 
Ray \& Bally, 2004; MR04). 
 
\subsection{Velocity of ``Micro\,-\,Jets''}
Out of the five ``micro\,-\,jets'' examined here HH\,220, HH\,158 and HH\,229 are optically visible in our images and their 
proper motions could be measured. We find velocities of 323 kms$^{-1}$ for HH\,220 (NW) and 251 kms$^{-1}$ for HH\,229 
(average of knots B,C, D and G) while HH\,158 (A, B and C) have an average velocity of 241 kms$^{-1}$ \citep{Eisloffel98}. 
These velocities are comparable to those found for jets from Class I sources, which are typically 200 -- 400 kms$^{-1}$  
\citep{Mundt87}. This suggests that the velocity of the jets remains high as their sources evolve, despite the fact that the
rate of accretion and ejection is about 10 -- 100 times smaller for Class II sources than Class I YSOs \citep{Hartigan95}.\\

Comparing our tangential velocity measurements for the distant objects to those of the ``micro\,-\,jets'' shows that the 
velocities are comparable despite the large difference in distance from their source. We cannot, of course, conclude that 
there is little or no deceleration in the outflow over parsec\,-\,scale distances as the ``events'' that give rise to the 
more distant HH objects could have originated from more violent (e.g. FUor) outbursts that those closer to the source.\\
 
\noindent There are two scenarios to be examined:\\
1. the ejection velocity at the source was much higher $\sim$ 10$^3$ years ago when the more distant objects were ejected 
and has decreased over the intervening years to its current velocity of 200 -- 300 kms$^{-1}$. Meanwhile the velocity of 
the older/more distant objects has decreased over time via interactions with the parent cloud, or\\ 

2. the velocity at the source has remained at approximately the same value over the last 10$^3$ years and the distant 
objects have not been slowed down much by their interaction with the ambient medium.\\

Large\,-\,scale numerical simulations are required to test whether deceleration might be important 
particularly in the context of the low density environment surrounding a CTTS. 

\subsection{Velocity Variations along the Outflows}
\label{sec-velocityvariations}
One of the aims of this proper motions study was to investigate how the velocity of HH objects varies with distance from 
the source. The pattern of velocity variation was not what we expected however -- we have found that the velocity appears to 
be constant despite the large distances between consecutive HH objects!\\

Proper motion studies done on the $\sim$ 3pc long HH\,34 outflow by \cite{Devine97} find a systematic decrease in proper 
motions with distance from the source. This was modelled by \cite{Cabrit00} who investigate whether this observed decrease 
is due to the ejection velocity at the source increasing over time or that the more distant objects have been slowed down 
by interactions as the propagate through the parent cloud. They find that the latter scenario is most likely in the case 
of HH\,34.\\

Our studies however do not show such a decrease in proper motion with distance for the CTTS-driven outflows, in fact they 
suggest that the velocity has remained approximately constant over at least 10$^3$ years. As the density of the ambient 
medium through which the outflow propagates is lower for Class II than Class I sources it is probable that the velocity of 
the more distant HH objects has not been greatly affected by its interactions with the parent cloud.\\ 

\section{Conclusions}
We have measured the proper motions of a number of HH objects associated with the CTTS-driven outflows from CW Tau, DG Tau, 
DO Tau, HV Tau C and RW Aur. The aims of this study were to firstly determine if the driving sources suggested in our 
original paper (MR04) are correct. Next was to find tangential velocities of the more distant HH objects in large\,-\,scale 
outflows, which are currently poorly known. Finally, we examine our results to see how tangential velocities evolve with 
distance from the source.\\ 

Our study confirms the previously suggested driving source for some of our objects: HH\,826 and HH\,827 are driven by CW 
Tau, HH\,702 by DG Tau and HH\,835 by RW Aur. However this study also reveals that some of the HH objects previously 
associated with these CTTSs are not actually driven by them - HH\,828 and HH\,829 are not driven by CW Tau, HH\,830 by DG 
Tau and HH\,831 and HH\,705 by either DO Tau or HV Tau C. These results are surprising in the case of HH\,831 and HH\,705 
which are well aligned with the ``micro\,-\,jets'' from DO Tau and HV Tau C respectively. While these results show that 
proper motions are necessary to conclusively determine the source of a HH object, we  contend that examination of the 
morphology and alignment of newly discovered HH objects can suggest candidate sources in the first instance, with proper 
motion studies undertaken at a later date when multi\,-\,epoch images are available.\\

While we now know that these CTTS-driven outflows are not as extended as previously thought in MR04, those from CW Tau, 
DG Tau and RW Aur still extend to $\sim$ 0.3pc -- 0.5pc from the source. The CW Tau and DG Tau outflows are mainly extended 
in the blue\,-\,shifted side, so it is quite likely that the receding red\,-\,shifted outflows are obscured. The 
blue\,-\,shifted outflows extend to the edge of the parent cloud (MR04), suggesting that these outflows have blown out. If 
they are now propagating into a lower density medium it is unlikely that we will observe them.\\
 
The tangential velocities determined for the more distant objects in the CW Tau, DG Tau and RW Aur outflows are typically 
200 kms$^{-1}$. How velocities of HH objects evolve with distance from the source is an important consideration in 
parsec\,-\,scale outflows. The proper motions of the ``micro\,-\,jets'' and HH objects close to the driving source show that
they have a typical velocity of the order of 200 kms$^{-1}$, with very little velocity variation over the lengths of these 
outflows. This suggests that major outbursts, roughly every thousand years, that give rise to giant HH complexes could 
primarily result from additional mass being deposited in an outflow rather than enormous increases in velocity. Further 
modeling is required to test this hypothesis.

\label{sec-conclusion}
\begin{acknowledgements}
D.\,Froebrich received support from
the Cosmo-Grid project, funded by the Program for Research in Third Level
Institutions under the National Development Plan and with assistance from the
European Regional Development Fund.
\end{acknowledgements}


\begin{thebibliography}{}
\bibitem[Anglada et al.(1992)]{Anglada92} Anglada, G., Rodr\'{\i}guez, L.~F., Canto, J., Estalella, R., \& Torrelles, J.~M.\ 
1992, \apj, 395, 494
\bibitem[Andre \& Montmerle(1994)]{Andre94} Andre, P., \& Montmerle, T.\ 1994, \apj, 420, 837 
\bibitem[Andre et al.(1993)]{Andre93} Andre, P., Ward-Thompson, D., \& Barsony, M.\ 1993, \apj, 406, 122
\bibitem[Andre et al.(2000)]{Andre00} Andre, P., Ward-Thompson, D., \& Barsony, M.\ 2000, Protostars and Planets IV, 59 
\bibitem[Bacciotti et al.(2002)]{Bacciotti02} Bacciotti, F., Ray, T.~P., Mundt, R., Eisl{\" o}ffel, J., \& Solf, J.\ 2002, 
\apj, 576, 222
\bibitem[Bally \& Devine(1997)]{Bally97} Bally, J., \& Devine, D.\ 1997, IAU Symp.~182: Herbig-Haro Flows and the Birth of 
Stars, 182, 29 
\bibitem[Barsony \& Kenyon(1992)]{Barsony92} Barsony, M., \& Kenyon, S.~J.\ 1992, \apjl, 384, L53
\bibitem[Bertin \& Arnouts(1996)]{Bertin96} Bertin, E.~\& Arnouts, S.\ 1996, \aas, 117, 393 
\bibitem[Bontemps et al.(1996)]{Bontemps96} Bontemps, S., Andre, P., Terebey, S., \& Cabrit, S.\ 1996, \aap, 311, 858 
\bibitem[Cabrit et al.(1990)]{Cabrit90} Cabrit, S., Edwards, S., Strom, S.~E., \& Strom, K.~M.\ 1990, \apj, 354, 687
\bibitem[Cabrit \& Raga(2000)]{Cabrit00} Cabrit, S., \& Raga, A.\ 2000, \aap, 354, 667 
\bibitem[Chini et al.(2001)]{Chini01} Chini, R., Ward-Thompson, D., Kirk, J.~M., Nielbock, M., Reipurth, B., \& Sievers, A.\ 
2001, \aap, 369, 155 
\bibitem[Devine et al.(1997)]{Devine97} Devine, D., Bally, J., Reipurth, B., \& Heathcote, S.\ 1997, \aj, 114, 2095 
\bibitem[Dougados et al.(2000)]{Dougados00} Dougados, C., Cabrit, S., Lavalley, C., \& M{\' e}nard, F.\ 2000, \aap, 357, L61
\bibitem[Duch{\^ e}ne et al.(1999)]{Duchene99} Duch{\^ e}ne, G., Monin, J.-L., Bouvier, J., \& M{\' e}nard, F.\ 1999, \aap, 
351, 954
\bibitem[Eiroa et al.(1994)]{Eiroa94} Eiroa, C., Miranda, L.~F., Anglada, G., Estalella, R., \& Torrelles, J.~M.\ 1994, \aap, 
283, 973 
\bibitem[Eisl{\" o}ffel \& Mundt(1992)]{Eisloffel92} Eisl{\" o}ffel, J., \& Mundt, R.\ 1992, \aap, 263, 292 
\bibitem[Eisl{\" o}ffel \& Mundt(1998)]{Eisloffel98} Eisl{\" o}ffel, J.~\& Mundt, R.\ 1998, \aj, 115, 1554
\bibitem[Elias(1978)]{Elias78} Elias, J.~H.\ 1978, \apj, 224, 857
\bibitem[Gahm et al.(1999)]{Gahm99} Gahm, G.~F., Petrov, P.~P., Duemmler, R., Gameiro, J.~F., \& Lago, M.~T.~V.~T.\ 1999, 
\aap, 352, L95
\bibitem[Gomez, Whitney, \& Kenyon(1997)]{Gomez97} Gomez, M., Whitney, B.~A., \& Kenyon, S.~J.\ 1997, \aj, 114, 1138 
\bibitem[Gomez de Castro(1993)]{GomezdeCastro93} Gomez de Castro, A.~I.\ 1993, \apjl, 412, L43
\bibitem[Hamann(1994)]{Hamann94} Hamann, F.\ 1994, \apjs, 93, 485
\bibitem[Hartigan, Edwards \& Ghandour(1995)]{Hartigan95} Hartigan, P., Edwards, S., \& Ghandour, L.\ 1995, \apj, 452, 736
\bibitem[Hartigan et al.(2001)]{Hartigan01} Hartigan, P., Morse, J.~A., Reipurth, B., Heathcote, S., \& Bally, J.\ 2001, 
\apjl, 559, L157
\bibitem[Herbig(1977)]{Herbig77} Herbig, G.~H.\ 1977, \apj, 217, 693
\bibitem[Hirth, Mundt, \& Solf(1994)]{Hirth94b} Hirth, G.~A., Mundt, R., \& Solf, J.\ 1994, \aap, 285, 929
\bibitem[Hirth, Mundt \& Solf(1997)]{Hirth97} Hirth, G.~A., Mundt, R., \& Solf, J.\ 1997, \aas, 126, 437
\bibitem[Hirth et al.(1994)]{Hirth94} Hirth, G.~A., Mundt, R., Solf, J., \& Ray, T.~P.\ 1994, \apjl, 427, L99
\bibitem[Itoh et al.(1999)]{Itoh99} Itoh, Y., Tamura, M., \& Gatley, I.\ 1996, \apjl, 465, L129 
\bibitem[Lada \& Wilking(1984)]{Lada84} Lada, C.~J.~ \& Wilking, B.~A.\ 1984, \apj, 287, 610
\bibitem[Lada(1987)]{Lada87} Lada, C.~J.\ 1987, IAU Symp.~115: Star Forming Regions, 115, 1
\bibitem[Mundt et al.(1987)]{Mundt87} Mundt, R., Brugel, E.~W., \& Buehrke, T.\ 1987, \apj, 319, 275 
\bibitem[Mundt \& Fried(1983)]{Mundt83} Mundt, R.~\& Fried, J.~W.\ 1983, \apjl, 274, L83
\bibitem[Mundt \& Eisl{\" o}ffel(1998)]{Mundt98} Mundt, R.~\& Eisl{\" o}ffel, J.\ 1998, \aj, 116, 860
\bibitem[McGroarty, Ray \& Bally(2004)]{McGroarty04a} McGroarty, F., Ray, T.~P., \& Bally, J.\ 2004, \aap, 415, 189
\bibitem[McGroarty \& Ray(2004)]{McGroarty04b} McGroarty, F., \& Ray, T.~P.\ 2004, \aap, 420, 975 {\bf MR04}         
\bibitem[Petrov et al.(2001)]{Petrov01} Petrov, P.~P., Gahm, G.~F., Gameiro, J.~F., Duemmler, R., Ilyin, I.~V., 
Laakkonen, T., Lago, M.~T.~V.~T., \& Tuominen, I.\ 2001, \aap, 369, 993
\bibitem[Ray(1987)]{Ray87} Ray, T.~P.\ 1987, \aap, 171, 145
\bibitem[Reipurth \& Bally(2001)]{Reipurth01} Reipurth, B., \& Bally, J.\ 2001, \araa, 39, 403 
\bibitem[Reipurth et al.(1997)]{Reipurth97} Reipurth, B., Bally, J., \& Devine, D.\ 1997, \aj, 114, 2708 
\bibitem[Reipurth et al.(2002)]{Reipurth02} Reipurth, B., Heathcote, S., Morse, J., Hartigan, P., \& Bally, J.\ 2002, \aj,
123, 362
\bibitem[Stapelfeldt et al.(2003)]{Stapelfeldt03} Stapelfeldt, K.~R., M{\' e}nard, F., Watson, A.~M., Krist, J.~E., 
Dougados, C., Padgett, D.~L., \& Brandner, W.\ 2003, \apj, 589, 410
\bibitem[Sun et al.(2003)]{Sun03} Sun, K., Yang, J., Luo, S., Wang, M., Deng, L., Zhou, X., \& Chen, J.\ 2003, Chinese 
Journal of Astronony and Astrophysics, 3, 458
\bibitem[Ward-Thompson(2002)]{Ward-Thompson02} Ward-Thompson, D.\ 2002, Science, 295, 76 
\bibitem[Wichmann et al.(1998)]{Wichmann98} Wichmann, R., Bastian, U., Krautter, J., Jankovics, I., \& Rucinski, S.~M.\ 
1998, \mnras, 301, L39


\end{thebibliography}
\end{document}